\documentclass[aps,pre,reprint]{revtex4-1}

\usepackage{amsmath}
\usepackage{amssymb}

\usepackage{graphicx}

\usepackage{color} 

\date{}

\usepackage{hyperref}

\begin{document}

\title{StochDecomp -  Matlab package for noise decomposition in stochastic biochemical systems}
\ \\
\begin{abstract}
{\noindent 
Tomasz Jetka$^{1}$, Agata Charzy\'nska\,$^{2}$, Anna Gambin\,$^{1}$, Michael P.H. Stumpf\,$^{3,*}$ \\
\noindent and Micha\l\  Komorowski$^{4,*}$ }
\ \\

\noindent $^{1}$Faculty of Mathematics Informatics and Mechanics,
University of Warsaw, Warsaw, Poland\\
$^{2}$Institute of Computer Science, Polish Academy of Sciences, Warsaw, Poland\\
$^{3}$Division of Molecular Biosciences, Imperial College London, London, United Kingdom\\
$^{4}$Institute of Fundamental Technological Research,
Polish Academy of Sciences, Warsaw, Poland\\
$^*$ Correspondence: \href{m.stumpf@imperial.ac.uk}{m.stumpf@imperial.ac.uk}, \href{m.komorowski@ippt.gov.pl}{m.komorowski@ippt.gov.pl}

\end{abstract}
%
%
%
%
%
%

\date{\today}

\pacs{}

\maketitle

\noindent {\bf Abstract \\}
\noindent Stochasticity is an indispensable aspect of biochemical processes at the cellular level. Studies on how the noise enters and propagates in biochemical systems provided us with nontrivial insights into the origins of stochasticity, in total however they constitute a patchwork of different theoretical analyses. Here we present a flexible and generally applicable noise decomposition tool,  that allows us to calculate contributions of individual reactions to the total variability of  a system's output. With the package it is therefore  possible to quantify how the noise enters and propagates in biochemical systems.  We also demonstrate and exemplify using the JAK-STAT signalling pathway that it is possible to infer noise contributions resulting from individual reactions directly from experimental data.  This is the first computational tool that allows to decompose noise into contributions resulting from individual reactions.\\
{\bf{Availability}}\\
\url{http://sourceforge.net/p/stochdecomp/}\\
{\bf{User Manual}}\\
\url{https://sourceforge.net/p/stochdecomp/wiki/}\\

\section{Introduction}
Although random fluctuations in molecule counts are inherent in biochemical signal processing their role is incompletely understood. The question which molecular species or parts of a network contribute most of the variability of a system or are responsible for most of the information loss has therefore attracted much attention in recent years. Numerous studies have analysed noise in signalling networks in detail and decomposed the noise into contributions attributable to fluctuations in mRNA and protein. Current software implementations offer a broad range of stochastic modeling methods  to analyse  stochastic properties of biochemical dynamics \citep{andrews2010detailed,ander2004smartcell,thomas2012intrinsic}. These tools however focus only indirectly on origins and propagation of stochasticty. To our knowledge, a software package to provide decomposition of noise into individual sources has been lacking. Recently we developed \citep{komorowski2013decomposing, wallace2013noise} a general method to analyse how the structure of biochemical networks gives rise to noise in its outputs.  In principle, this allows us to efficiently calculate the contribution each reaction makes to the variability in all concentrations for any modelled network. One can then ask quantitatively how changes in reaction rates, molecular concentrations, or even in network structure affect the variability in an output of interest.  Moreover, if experimental data is available and a posterior of model parameters can be generated  the contribution of individual reactions can be estimated from data. To our knowledge our approach \citep{komorowski2013decomposing}  was the first method which enabled assignment of origins of variability to individual reactions and arbitrarily defined network components.\\
Below we provide a general description of the package. Details are presented in the {\emph{ User Manual}}, which includes theoretical foundations of the method, user manual and examples. In a comprehensive analysis of the JAK-STAT signalling pathway we infer individual contributions from experimental data published in  \citep{swameye2003identification}.
\vspace{-5mm}
\section{Methods}
The linear noise approximation (LNA) was employed to model stochastic chemical kinetics \citep{ja_LNA}. It represents the covariance $\Sigma$, a matrix quantifying the noise in every network component, in form of ordinary differential equations (ODEs) 
\begin{equation}\label{Sigma}
\frac{d\Sigma}{dt}= A(t)\Sigma+\Sigma A(t)^{T}+D(t)
\end{equation}
Because (\ref{Sigma}) is linear in $\Sigma$, and $D$ decomposes into a sum across reactions,  $\Sigma$ likewise decomposes into a sum across reactions \citep{komorowski2013decomposing}
\begin{equation}
\Sigma=\Sigma^{(1)}+...+\Sigma^{(r)},
\end{equation}
where $r$ denotes the number of reactions in the system.  From a specification of the network, we calculate the response matrix A, which describes how the network state instantaneously responds to fluctuations, and the dissipation matrix $D$, which describes the contribution of count noise. This enables us to identify the origins of cell-to-cell variability in dynamical biochemical systems and pinpoint, if warranted, individual reactions.\\

\noindent {\bf{Implementation.}} 
The package is implemented as a set of Matlab functions. Any model to be analysed using these functions needs to be defined in terms of a stoichiometry matrix,  a Matlab function containing reaction rates, and a vector of parameter values. From these files a set of ODEs is generated using the Matlab Symbolic Toolbox, which is required to run the package. Equations are then solved using the Matlab ODE solver and  solutions provide the variance decomposition.  Functions to provide a graphical output has been implemented to visualise decomposition both in- and out-of-steady-state.\\

\begin{figure}[htb]
\begin{center}
\includegraphics[scale=0.205]{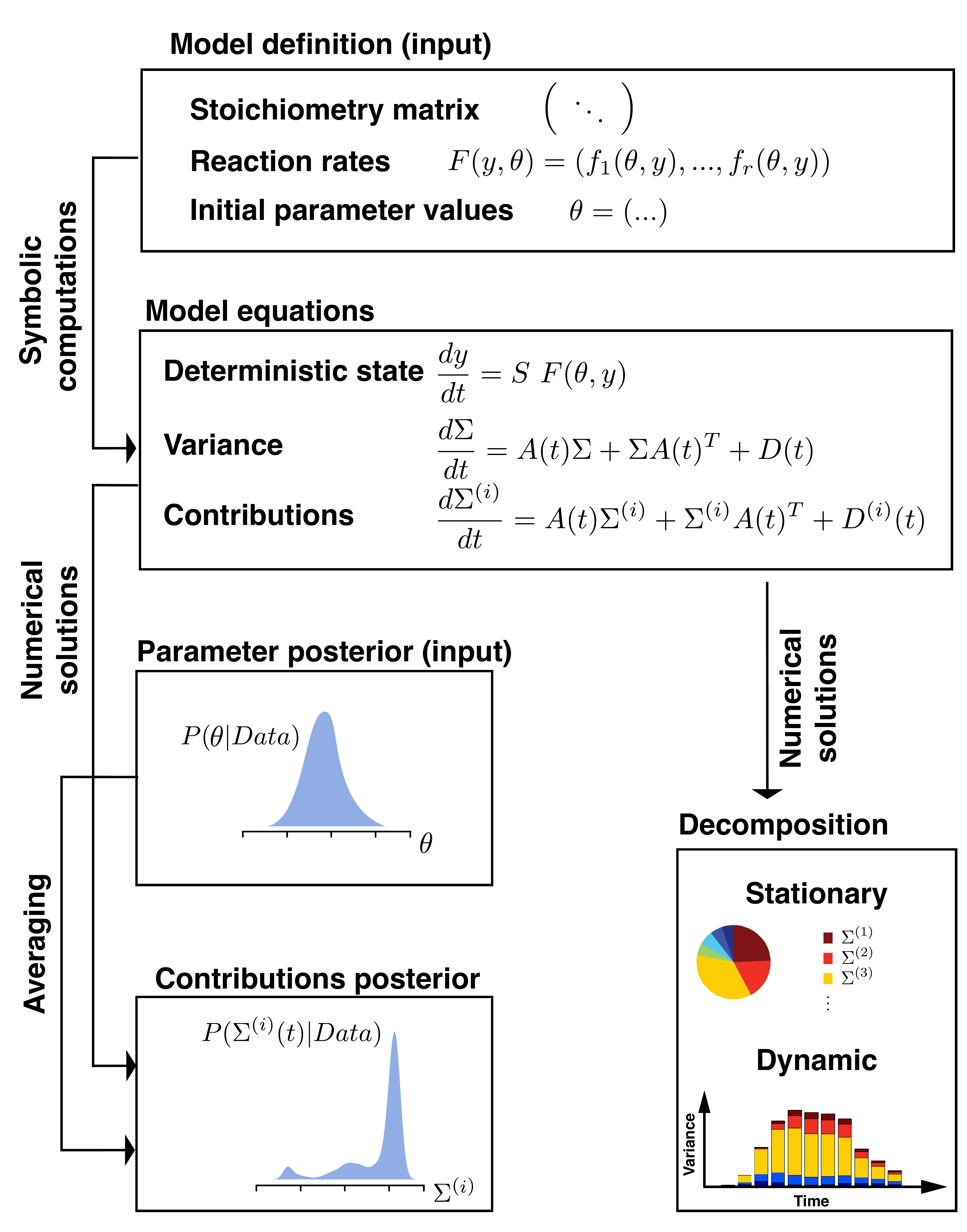}
\vspace{-5mm}
\end{center}
\caption{Input-output flow chart of the StochDecomp package. The stoichiometry matrix, reaction rates and initial parameter guesses are translated by means of the symbolic computations into a set of ODEs. Stored ODEs are then used to numerically analyse the model, particularly decompose the variability into contributions resulting from each of the model reactions either in steady state or out-of-steady-state. If in addition experiential data together with a posterior distribution on model parameters are available the posterior distribution on individual contributions can be calculated. \label{IO}}
\end{figure}

\noindent {\bf Applicability.}
The applicability of the package is limited by the validity of LNA, which is discussed in details in \citep{wallace2012linear, 
komorowski2013decomposing}. In principle, the tool allows us to take any modelled network, and efficiently calculate the contribution each reaction makes to the variability in all concentrations.  Specifically, the functions provided allow to: (i) symbolically generate ODEs describing the system and individual reaction contributions;  (ii) numerically compute variance decomposition and visualise obtained results; (iii) infer contributions of individual reactions from experimental data if posterior distribution is provided. The flow chart describing input-output relationship of the package is presented in Fig. \ref{IO}. 
\ \\

\noindent {\bf{Biological relevance.}}
Using experimental data of \citep{swameye2003identification} we infer the sources of variability in the JAK-STAT signalling pathway. Firstly we employed the PUA Matlab package \citep{vanlier2012integrated} to generate posterior distribution of model parameters as described in \citep{vanlier2012integrated}. Secondly the posterior was  used as an input for our tool. A comprehensive analysis of the dynamic variability in the nuclear concentration of  STAT complexes, which is a factor activating a downstream response, reveals that: (i) In the absence of extrinsic noise, the fluctuations in the number of nuclear complexes originate largely from trafficking of the complexes into the nucleus. (ii) In the presence of the extrinsic noise, understood as fluctuations in Epo concentration, the network acts as a low pass filter. The extrinsic noise is major source of variability if the fluctuations in Epo concentration are slow; (iii) The overall variability of the nuclear concentration of  STAT complexes is not sensitive to parameters. Contributions of certain reactions, however, are sensitive and change by an order of magnitudes for the parameters within the posterior. The details of our findings are described in the section 3 of the {\emph{User Manual}}. 
\hspace{-50mm}
\section{Discussion}
StochDecomp is a novel, computationally efficient and extendable Matlab package for computational analysis of noise origin in the stochastic models of biochemical reactions. It implements a general method to analyse how structure of a biochemical network gives rise to noise in its outputs. It can be applied to virtually any biochemical system and its applicability is only limited by the usage of the LNA. If experimental data on studied system is available, the package in combination with Bayesian inference software e.g. PUA  \citep{vanlier2012integrated} can be used to infer the variance contributions of individual reactions directly from experimental data. This makes our development unique and complementary to existing implementations.
\hspace{-5mm}
\section*{Acknowledgement}
{
\noindent TJ and MK were supported by the Foundation for
Polish Science (HOMING 2011-3/4);
AC by research fellowship within {\it Information technologies: research
and their interdisciplinary applications} (POKL.04.01.01-00-051/10-00);
AG by National Science Center (2011/01/B/NZ2/00864);
MPHS by BBSRC (BB/G020434/1). MPHS is a Royal Society Wolfson Research Merit Award holder.  MK is EMBO Installation Grantee.
}

\end{document}